\documentclass[10pt,twocolumn,twoside]{IEEEtran}

\usepackage{amsmath,amssymb,bbm}
\usepackage{color,graphicx}

\usepackage{array}
\usepackage{mdwmath}
\usepackage{mdwtab}
\usepackage{eqparbox}

\def\pP{{\mathbb P}}

\def\eE{{\mathbb E}}
\def\errorprob{{e}}

\newtheorem{example}{Example}[section]
\newtheorem{theorem}{Theorem}[section]

\def\proof{\noindent\textit{Proof:} }
\def\endproof{{\hfill $\clubsuit$ \medskip}}

\begin{document}

\title{Bandlimited Field Reconstruction from Samples Obtained at
Unknown Random Locations on a Grid}

\author{\IEEEauthorblockN{Ankur Mallick and Animesh Kumar\\
Department of Electrical Engineering\\
Indian Institute of Technology Bombay\\
Mumbai, India 400076 \\
Email: ankur.mallick24@gmail.com, animesh@ee.iitb.ac.in}}

\maketitle

\begin{abstract}

We study the sampling of spatial fields using sensors that are location-unaware
but deployed according to a known statistical distribution. It has been shown
that uniformly distributed location-unaware sensors cannot infer bandlimited
fields due to the symmetry and shift-invariance of the field.

This work studies asymmetric (nonuniform) distributions on location-unaware
sensors that will enable bandlimited field inference. For the sake of analytical
tractability, location-unaware sensors are restricted to a \textit{discrete
grid}. Oversampling followed by clustering of the samples using the probability
distribution that governs sensor placement on the grid is used to infer the
field . Based on this clustering algorithm, the main result of this work is to
find the \textit{optimal} probability distribution on sensor locations that
minimizes the detection error-probability of the underlying spatial field.  The
proposed clustering algorithm is also extended to include the case of signal
reconstruction in the presence of sensor noise by treating the distribution of
the noisy samples as a mixture model and using clustering to estimate the
mixture model parameters.

\end{abstract}

\section{Introduction}
\label{sec:intro}

Distributed sensing of spatial fields using an array of distributed low-power
sensors is an active area of interest. In the past, sampling techniques have
been proposed as one method of distributed sensing, where the spatial field can
be modeled as a distributed
signal~\cite{kumarIRD2010,kumarIRH2011,murraybruceDE2015,unnikrishnanVS2013,marcodlnoIPSN2003,ranieriVS2013,sharmaKS2015}.
In all these works and many more, a central theme is that the location of
sensors is known. In practice, the location of sensor can be ascertained by
localization algorithms~\cite{patwariAKHMCL2005,wymeerschLWC2009}, or by using
extra equipment such as a GPS receiver (global positioning system receiver). These
techniques will have added cost in terms of power used and hardware required.
An alternate option to having expensive sensors or expensive localization
algorithms is to work with sensors which are location unaware.

Recently, bandlimited field estimation with location unaware sensors in a
distributed setup has been studied~\cite{kumarO2015,kumarB2016}.  This is an
interesting paradigm where the key idea is to utilize a multitude of location
unaware sensors (oversampling) and leverage the random distribution on their
spatial locations to reconstruct the spatial field. In a negative result, it is
known that uniformly distributed location unaware sensors do not infer the field
uniquely up to a shift and a flip of the underlying signal~\cite{kumarO2015}.
The negative result has its root in the symmetry of a uniform distribution and
the shift-invariance properties of bandlimited fields. In this work, it will be
further noted that the negative result extends to the spatially scaled versions
of a bandlimited field as well. For example, spatial fields $g(x)$ and $g(2 x)$
will be indistinguishable by readings observed from uniformly distributed
location unaware sensors.  These negative results motivate the study of an
asymmetric (statistical) distribution of location unaware sensors, that may
enable bandlimited field reconstruction. 

A bandlimited field is a nonlinear function of the location.  If the location of
each sensor is random and unknown, then a bandlimited field operating on this
randomness is observed through samples.  This process is nonlinear and leads to
difficult inference problems.  To facilitate analysis, in this first exposition
on the topic, the location of the sensors is restricted to a random point on an
\textit{equi-spaced discrete grid}.\footnote{This may arise in scenarios where
location information is masked to preserve the identity of the sensors, or to
reduce the amount of data that needs to be transmitted.} The location
unawareness of sensors will be overcome by using oversampling in our setup.  For
field reconstruction purposes, a technique is required which associates the
samples of the field to their respective locations on the equi-spaced
deterministic grid in the sampling interval of interest. One such technique is
sample clustering as explained next.

With oversampling, samples obtained from sensors can be clustered together to
infer which sample belongs to which spatial location on the equi-spaced grid
where the sensors are present. The key insight behind our clustering method is
as follows. All the sensors present at a location will record the same field
value. The probability with which a sensor falls at any location in the
deterministic grid is a parameter of choice.  If $p$ is the probability with
which a sensor falls at a given location, then $\approx np$ will be the number
of samples obtained from there, as $n$ (total number of samples across all
locations) becomes large. If the sensors have unequal probability of falling at
different locations, the locations of samples can be \textit{detected} by using
their expected frequency.  The success of this clustering scheme will depend on
the probability distribution that governs sensor placement on the grid.  One of
the \textit{main results} of this work is to find the \textit{optimal}
probability distribution on sensor locations that minimizes the detection
error-probability of the underlying spatial field (see
Section~\ref{sec:performance}). 

Since samples obtained from real world signals are typically affected by
measurement noise, the effect of additive noise on samples is also explored on
our clustering algorithm and associated field reconstruction. A different
clustering approach, that banks upon methods used in machine learning, is used
to associate the various field samples with their respective locations. Then,
the observed noise-affected field samples are used to estimate the field values.
The distribution of the noise-affected samples is modeled by a mixture model and
the special case of Gaussian noise is analysed to show that our approach works
fairly well in most cases even in the presence of noise. Simulation results are
presented to verify our algorithm  (see Section~\ref{sec:noisy}).

\textit{Prior art:} Estimation of bandlimited fields from samples taken at
unknown but statistically distributed sampling locations was studied by
Kumar~\cite{kumarO2015,kumarB2016}.  Reconstruction of discrete-time bandlimited
fields from unknown sampling locations was studied by Marziliano and
Vetterli~\cite{marzilianoV2000} in a combinatorial setting.  Estimation of
periodic bandlimited signals with random sampling locations has been studied by
Nordio~et~al.~\cite{nordioCVP2008}, where the locations are obtained by a
perturbation of the deterministic equi-spaced grid.  Their work is related to
estimation or ``denoising'' of bandlimited fields.  In this work, sampling of
physical fields is addressed via oversampling when sensor locations are
restricted to a deterministic equi-spaced grid. Sensors are location unaware,
and a clustering algorithm will be used to associate field samples with their
respective locations.  The design of probability distribution on sensor
locations is addressed in this work to minimize the probability of incorrect
association of field samples with the sampling locations on the equi-spaced
grid.  Finally, a clustering algorithm is presented in this work to tackle
measurement noise at the sensors.

\textit{Notation:} Space will be denoted by $x$. Spatial fields will be denoted
by $g(x)$ and its variants, and the Fourier series coefficients will be denoted
by $a[k]$ and its variants.  
Probability distribution function (pdf) of a random variable $Y$ will be denoted
by $f_{Y}(y)$.  A Gaussian distribution with mean $\mu$ and standard deviation
$\sigma^2$ will be denoted by ${\cal N}(\mu, \sigma^2)$, while $j = \sqrt{-1}$
will be the imaginary root of $-1$. The probability and expectation operators
will be denoted by $\mathbb{P}$ and $\mathbb{E}$ respectively. All vectors are
column-vectors. 

\textit{Organization:} The sampling system model and the reconstruction
methodology are presented in Section~\ref{sec:setup}. In
Section~\ref{sec:performance}, the minimization of field detection
error-probability is addressed. Throughout the analysis in this section it is
assumed that the samples are not corrupted by any noise. In
Section~\ref{sec:noisy}, the algorithm for field estimation from
measurement-noise affected samples is addressed.  Finally, conclusions are
presented in Section~\ref{sec:conclusions}.

\section{System model, reconstruction, and related results}

\label{sec:setup}

The spatial field model and its properties, the sampling model used, and related
theoretical results are reviewed in this section in this section. The field
model appears first.

\subsection{Spatial field model}
\label{sec:spatfield}

The spatial field varies with space and time. A distributed array of sensors
will be used to acquire the field, so the following model on the field will be
applicable to each time snapshot of the field.  Time snapshots of spatially one
dimensional fields are considered. The spatial field $g(x)$ is assumed to be
real valued, bounded, and have a finite spatial support in $[0,1]$, It is
assumed that the field has a Fourier series with finite number of terms, that
is, 
\begin{align}
g(x) = \sum_{k = -b}^{b} a[k] \exp(j 2\pi k x) \label{eq:fseries}
\end{align}
where $0 \leq x \leq 1$, and where $a[k]$ are the Fourier series coefficients of
$g(x)$ and $b$ is a \textit{known} bandwidth parameter.  

For simplicity of notation, define $s_b := 1/(2b+1)$ as a grid spacing 
parameter and
$\phi_k := \exp(j 2 \pi k s_b), -b \leq k \leq b$.  Let $\Phi_b$ be defined as
\begin{align}
\Phi_b
= \left[ \begin{array}{c c c} 1 &  \ldots & 1 \\ 
\phi_{-b} & \ldots  & \phi_{b} \\
\vdots &  &  \vdots \\ 
(\phi_{-b})^{2b} & \ldots  & (\phi_{b})^{2b} \end{array} \right].  \nonumber
\end{align}
The columns of $\Phi_b$ are orthogonal and a sampling theorem ensures
that~\cite{nordioCVP2008,oppenheimsyd1999}:
\begin{align}
\vec{a} = (\Phi_b)^{-1} \vec{g} = \frac{1}{(2b + 1)} \Phi_b^\dagger \vec{g},
\label{eq:afromg}
\end{align}
where $\vec{a} = (a[-b], a[-b+1], \ldots, a[b])^T$, where $\Phi_b^\dagger$ is
the conjugate transpose of $\Phi_b$, and $\vec{g} = \left( g(0), g(s_b), \ldots,
g(2b s_b)\right)^T$.  From~\eqref{eq:afromg}, $\vec{a}$ and $g(x)$ can be
obtained using the samples in $\vec{g}$.

It will also be assumed that $g(is_b)$ are \textit{distinct} for different
values of $0 \leq i \leq 2b$.  This feature will be useful in associating
samples with their locations.  The unequal values of the field at grid locations
can be justified if the corresponding Fourier series coefficients are linearly
independent continuous random variables.  Let $\vec{a}$ be the realization of a
linearly independent continuous random vector (as may be the case for Fourier
coefficients of naturally occurring fields). If $g(m s_b) = g(n s_b)$, then 
\begin{align}
\sum_{k=-b}^{b} a[k] (\exp(j2 \pi kms_{b})-\exp(j2 \pi kns_{b}))=0.
\end{align}
This condition will be never satisfied, since $\vec{a}$ is a linearly
independent continuous random vector, and therefore its linear combination is
non-zero almost surely.

\subsection{Sensor deployment model}
\label{sec:sensordeploy}

\begin{figure}
\centering
\includegraphics[scale=0.5]{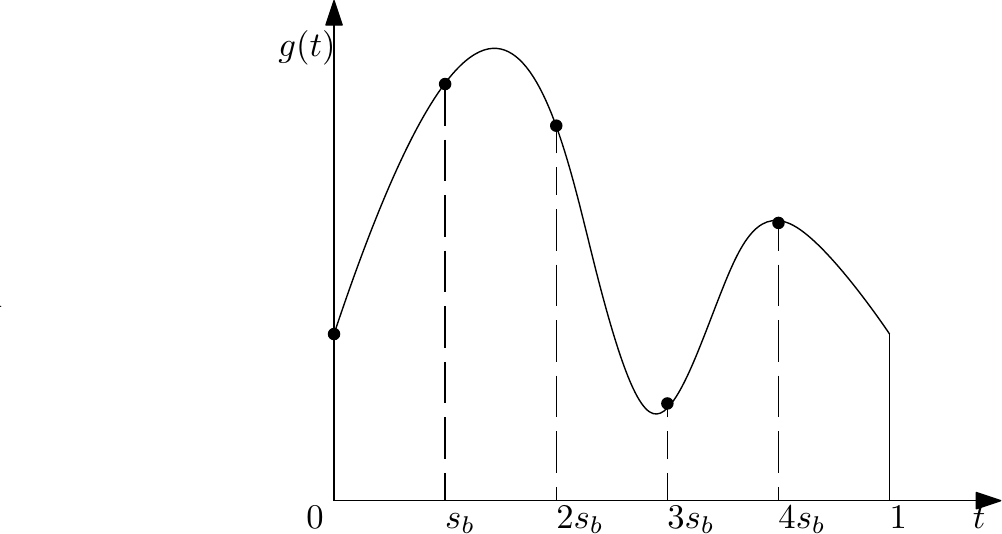}
\caption{\label{fig:samplemodel} Sampling model for a signal $g(t)$ with $b=2$
i.e. $s_{b}=1/5$.}
\end{figure}

A discrete-valued non-uniform distribution is considered for bandlimited field
inference.  It will be assumed that a sensor is at location $X$ such that $X = i
s_b$ with probability $p_i$ where $i = 0, 1, \ldots, 2b$ and $\sum_{i=0}^{2b}
p_{i}=1$.  Correspondingly,
\begin{align}
g(X) = g(i s_b) \mbox{ with probability } p_i, \quad i = 0, 1, \ldots, 2b
\label{eq:locmodel}
\end{align}
since it is assumed that $g(0), g(s_b), \ldots, g(2b s_b)$ are distinct.  In our
model (illustrated in Fig.~\ref{fig:samplemodel}), the sensor falls at $i s_b, 0
\leq i \leq 2b$ but its location, that is the index $i$, is \textit{not known}.
The parameter $\vec{p} := p_0, p_1, \ldots, p_{2b}$ will be treated as a
\textit{design choice} to optimize a  performance criterion of choice (see
Section~\ref{sec:performance}).  It will be assumed that elements of $\vec{p}$
are distinct (to break symmetry in the distribution of sensor-locations).
Without loss of generality, it will be assumed that
\begin{align}
p_0 < p_1 < \ldots < p_{2b}. \label{eq:porder}
\end{align}
In our sampling model, i.i.d.~samples $g(X_1), g(X_2), \ldots, g(X_n)$ are
available for the detection of spatial field, where $n$ corresponds to
oversampling.\footnote{It is desirable to address the setup where each sensor's
location $X$ is realized from an asymmetric continuous distribution supported in
$[0,1]$. In the limit of large number of measurements, the probability
distribution of $g(X)$ random variable will be available. Since $g(x)$ is a
non-linear function with oscillatory behavior, obtaining the Fourier series of
$g(x)$ from the statistics of $g(X)$ is nonlinear and analytically difficult.}

\subsection{Difficulty in field reconstruction with samples at uniformly
distributed locations}

To motivate the use of an asymmetric distribution on the sampling locations, we
first consider the case where the unknown sampling locations are realized
according to a uniform distribution, that is, $X \sim \mbox{Uniform}[0,1]$. Let
$U_1, U_2, \ldots, U_n$ be the (random) sampling locations according to this
uniform distribution; then, the corresponding sampled field values are
$g(U_{1}), g(U_{2}),... g(U_{n})$. For this section, also assume that $g(x)$ is
bounded in amplitude. Without loss of generality, let $|g(x)| \leq 1$ for all $x
\in [0,1]$. The empirical cumulative distribution function
\begin{align} 
F_{g,n}(\theta) = \frac{1}{n}\sum\limits_{i=1}^{n} \mathbbm{1}(g(U_{i}) \leq
\theta) \label{eq:glivenko}
\end{align}
for $\theta \in [-1,1]$ completely characterizes the field values $g(U_{1}),
g(U_{2}),... g(U_{n})$ (up to a permutation) and vice-versa.  By the
Glivenko-Cantelli theorem, for every $\theta \in [-1,1]$, $F_{g,n}(\theta)$ in
\eqref{eq:glivenko} converges almost surely to $\pP(g(U)\leq \theta)$. For a
uniform random variable, $\pP(g(U) \leq \theta)$ corresponds to length of the
level set $\{u: g(u) \leq \theta\}$. It has been shown that a bandlimited field
when shifted or space-reversed results in the same length of level
set~\cite{kumarO2015}. That is,
\begin{align}
\pP(g(U) \leq \theta) = \pP(g_1(U) \leq \theta) = \pP(g_2(U) \leq \theta)
\label{eq:ambiguity}
\end{align}
where $g_1(t) = g(t - s)$ and $g_2(t) = g(s - t)$ for any $s \in [0,1]$ and any
$\theta \in [-1,1]$.  This means by observing the distribution $\pP(g(U) \leq
\theta), \theta \in [-1,1]$, the field $g(x)$ cannot be inferred due to
ambiguity in phase and direction.  However, it is not clear if shift and
direction are the only ambiguities in the estimation of the field. In other
words, is it possible to claim that the field can be obtained up to a delay and
direction ambiguity as hinted in \eqref{eq:ambiguity}?  It is shown next that
\textit{scale ambiguity} is also present and this makes sampling a spatial field
with uniformly distributed sensors impossible. 

Let $g(x)$ be a field with bandwidth $2\pi$, which is less than $2b\pi$.
Consider the field $g_{3}(t)=g(mt)$ for any positive integer $m \leq b$.  Then
$g_3(t)$ is bandlimited with bandwidth up to $2 b \pi$. It will be shown that 
\begin{align}
\pP(g(U) \leq \theta) = \pP(g_3(U) \leq \theta) \label{eq:scaleambi}
\end{align}
for any $\theta \in [-1, 1]$ and any $1 \leq m \leq b$.  The core idea behind
the proof is the accounting of the length of level set.  Let 
\begin{align}
\{u: g(u)\leq \theta\} = [x_{0}, x_{1}] \cup \ldots & \cup [x_{N-1}, x_{N}] \cup
\nonumber \\
& \{x_{N+1}, \ldots, x_{M}\} \label{eq:lset1}
\end{align}
where $x_0, x_1, \ldots, x_{M}$ depend on $g(u)$ and $\theta$.  Then, for $m =
2$,
\begin{align}
\{&u: g_3(u) \leq \theta\} \\
& = \{u: g(2u)\leq \theta\} \\
& = \left[ \frac{x_{0}}{2}, \frac{x_{1}}{2}\right] \cup \ldots \cup \left[
\frac{x_{N-1}}{2}, \frac{x_{N}}{2}\right] \cup  \nonumber \\
& \mbox{\hspace{0.5cm}} \left[ \frac{x_{0}+1}{2},
\frac{x_{1}+1}{2}\right] \cup \ldots \cup \left[ \frac{x_{N-1}+1}{2},
\frac{x_{N}+1}{2}\right] \cup \nonumber \\
& \mbox{\hspace{0.5cm}} \left\{ \frac{x_{N+1}}{2}, \ldots, \frac{x_{M}}{2},
\frac{x_{N+1} + 1}{2}, \ldots, \frac{x_{M} + 1}{2} \right\}. \label{eq:lset2}
\end{align}
Observe that the lengths of the level sets in \eqref{eq:lset1} and
\eqref{eq:lset2} are equal to 
\begin{align}
(x_1 - x_0) + (x_3 - x_2) + \ldots + (x_N - x_{N-1})
\end{align}
with the note that single points $x_{N+1}, \ldots, x_{N+M}$ in the level sets
are zero in length (measure). This observation is true for any $g(x)$ with $3$
Fourier series coefficients, and any $x \in [-1,1]$. So $g(x)$ and $g(2x)$ have
the same length of level sets and consequently same distribution $\pP(g(U) \leq
\theta)$.  This result can be also shown in a similar manner for $m = 3, \ldots,
b$.  Thus, even if $n \rightarrow \infty$, the field $g(x)$ cannot be inferred
uniquely from $F_{g,n}(\theta)$ which converges to $P(g(U)\leq \theta)$, $\theta
\in [-1,1]$. This result strengthens our claim that \textit{an alternate
distribution} is needed on the unknown sampling locations.

\subsection{Sanov's theorem and hypothesis testing limits}

It is natural to use likelihood ratio test in a detection setup.  To analyze the
detection error-probability, large deviation analysis setup will be used.
Sanov's theorem, which addresses the asymptotic likelihood properties with
respect to an incorrect probability model, will be
used~\cite[Chap~11.4]{coverte1991}.  Let $R_{1},\ldots, R_{n}$ be i.i.d.~random
variables with discrete distribution $\vec{p}$. Then, the observed distribution
of $R_1, \ldots, R_n$ lies in the closed set $E$ with the following probability
\begin{align}
\lim_{n\to\infty} \frac{1}{n}\log_{2} \left[ \pP(R_1^n \in E) \right]
=-D(\vec{q}{*}\parallel \vec{p} ) \label{eq:sanov}
\end{align}
where $\vec{q}{*}=\arg\min_{\vec{q} \in E} D(\vec{q} \parallel \vec{p})$ is the
distribution in $E$ that is the closest to $\vec{p}$ in the Kullback Leibler
divergence or relative entropy terms. In other words as $n$ becomes large
\begin{align}
\pP( R_1^n \in E) \propto 2^{- n D(\vec{q}{*}\parallel \vec{p}) }.
\end{align}
The quantity $D(\vec{q}* \parallel \vec{p})$ will be termed as the
\textit{error-exponent} in this work.

\section{Field detection and its performance}
\label{sec:performance}

In this section, our field detection algorithm and its performance is discussed
where the samples are not affected by measurement noise. The optimal probability
distribution, with which sensors should be deployed at various sampling
locations, will be derived to minimize the detection error probability of the
underlying spatially bandlimited field in this section.

\subsection{The field detection algorithm}

\label{sec:detectalgo}

At first, it will be shown that the type of observed field values is a
sufficient statistic for detection. Then, a clustering algorithm will be
presented that uses the type of observed field values.

Since the field is sampled at $(2b+1)$ distinct locations, the samples $Y_1,
\ldots, Y_n$ take $(2b+1)$ distinct values as discussed in
Section~\ref{sec:spatfield}. Let $\vec{V} = [V_0, V_1, \ldots, V_{2b}]^T$ denote
the vector of these values. It is observed that field values at the sampling
locations $g(k s_b), 0 \leq k \leq 2b$ are an unknown permutation of the
elements of $\vec{V}$. There are $(2b+1)!$ distinct permutations of $\vec{V}$.
The goal of our field detection algorithm should be to assign the correct
location $k s_b, 0 \leq k \leq 2b$ to each element of $\vec{V}$. That is, the
correct permutation which relates $\vec{V}$ with the true field samples is
desired.

Let $\vec{M} = [M_0, \ldots, M_{2b}]^T$ denote the vector of types corresponding
to the values in $\vec{V}$ (i.e., $M_k$ is the type of value $V_k$). Each
permutation of the values corresponds to a permutation of the types as well. Let
$\rho$ be a permutation, and let $(\vec{V}^\rho, \vec{M}^\rho)$ be the permuted
versions of the value and type vectors, respectively. The goal of detection is
to identify the correct permutation $\rho^*$ that leads to the correct
assignment of values to the locations.

The field is sampled according to a distribution $\vec{p}$ as defined in
Section~\ref{sec:sensordeploy}. Since $g(ks_b), 0 \leq k \leq 2b$ are a
permutation of $\vec{V}$, so the probability distribution of observations $Y_1,
\ldots, Y_n$ is given by
\begin{align}
f(\vec{Y}|\rho) = \prod_{k = 0}^{2b} p_k^{M_k^\rho}.
\end{align}
This satisfies the Fisher-Neyman factorization theorem for sufficient
statistic~\cite{bickelDM2001} and $\vec{M}$ is the sufficient statistic for our
problem. 

Based on the readings $g(X_1), g(X_2), \ldots, g(X_n)$, the field $g(x)$ has to
be detected. The correct detection of the field corresponds to the correct
association of recorded field values with their respective locations.
From~\eqref{eq:porder} and Section~\ref{sec:spatfield}, $\{g(i s_b), p_i\}$
pairs are distinct in both the elements. Each sensor gets deployed at $X = i
s_b$, and subsequently records $g(is_b)$, with probability $p_i$.  The following
\textit{clustering algorithm} will be used to ascertain the field samples
$g(is_b)$ that specify the entire field $g(t)$ (see~\eqref{eq:afromg}):
\begin{enumerate}
\item The readings $Y_1 := g(X_1), \ldots, Y_n := g(X_n)$, with $X_i$ unknown
and in the set  $\{0, s_b, \ldots, 2b s_b\}$, are collected.
\item The values $Y_1, Y_2, \ldots, Y_n$ are clustered into (\textit{value},
\textit{type}) pairs.  Equal values (\textit{value}) in $Y_1, Y_2, \ldots, Y_n$
are grouped together and the number of equal values (\textit{type}) is recorded.
At most $2b+1$ such pairs exist since $2b+1$ field values are sampled to
generate $Y_1, \ldots, Y_n$.
\item Empirical probabilities \textit{type}$/n$ for each \textit{value} are
calculated. For large $n$, by weak law of large numbers, the empirical
probability \textit{type}$/n$ of each \textit{value} will be ``near'' the
correct $p_i$ in $\vec{p}$.
\item The \textit{value} with smallest empirical probability is assigned to
$g(0)$, the \textit{value} with next smallest empirical probability is assigned
to $g(s_b)$, and so on till $g(2b s_b)$.
\end{enumerate}
The clustering algorithm above, with a correct and an incorrect detection of the
field, is illustrated by an example below.
\begin{example}[Detection of field with location unaware sensors on a grid]
Consider a signal $g_1(t)$ with bandwidth parameter $b = 1$, and $s_b =
\frac{1}{2b+1} = \frac{1}{3}$.  The correct field samples at $0, s_b, 2s_b$ are
$g_1(0)=1.06, g_1(1/3)=1.80, g_1(2/3)=0.14$.

The field is sampled using $n = 10$ randomly realized values of sensor's
location in the set $\{0, 1/3, 2/3\}$.  In one realization of this statistical
experiment, the  $10$ observed samples are $1.80, 0.14, 0.14, 1.06, 1.80,$ $0.14,
1.80$, $1.06, 0.14, 0.14$. The (\textit{value}, \textit{type}) pairs are $(1.06,
2)$, $(1.80, 3)$, and $(0.14, 5)$. The above clustering algorithm concludes that $g_1(0) =
1.06$, $g_1(1/3) = 1.80$, $g_1(2/3) = 0.14$, and it correctly associates the
field values to the sample locations in this case.

In another realization of the same statistical experiment, 
the $10$ observed samples are $1.06, 0.14, 0.14,$ $1.06,1.80, 0.14, 1.80, 1.06$,
$0.14, 0.14$. The (\textit{value}, \textit{type}) pairs are $(1.06, 3)$, $(1.80,
2)$, and $(0.14, 5)$. The above clustering algorithm concludes that $g_1(0) =
1.80$, $g_1(1/3) = 1.06$, $g_1(2/3) = 0.14$, and it incorrectly associated the
field values to the sample locations in this case.  \hfill $\clubsuit$
\end{example}
Next, the exponent of the detection error probability of the clustering
algorithm will be minimized as $n \rightarrow \infty$.  For further discussions
on the clustering algorithm described above, define
\begin{align}
N_i := \sum_{j= 1}^n \mathbbm{1} \left[ Y_j = g(i s_b) \right]
\end{align}
as the type random variable of $g(i s_b)$ in $n$ field observations. Note that
$\eE (N_i) = n p_i$ for $0 \leq i \leq 2b$.  Since $0 < p_0 < p_1 < \ldots <
p_{2b}$, as $n \rightarrow \infty$ in the above-mentioned clustering algorithm,
by the weak law of large numbers it is expected that
\begin{align}
0 < N_0 < N_1 < \ldots < N_{2b} \label{eq:typecond}
\end{align}
with high-probability.  If the statistical event in~\eqref{eq:typecond} on type
random variables is not true, the above-mentioned clustering algorithm will
result in erroneous field detection. The probability of correct detection
in~\eqref{eq:typecond} will be maximized by choosing the sensor deployment
distribution $\vec{p}$ in what follows. 

\subsection{Field detection error-probability minimization}
\label{sec:probmin}

The spatial field is detected correctly when the condition
in~\eqref{eq:typecond} is satisfied. Let $\errorprob_n$ be the detection
error-probability. The error-exponent, as the number of sensors $n$ gets large,
in the detection error-probability $\errorprob_n$ will be maximized in this section. Note
that,
\begin{align}
\errorprob_n & =  \pP \big[ (0<N_{0}<N_{1}<...<N_{2b})^{c} \big] \label{eq:pe} \\
& =  \pP \big[ \{ N_{0}=0 \} \cup \{ N_{0}\geq N_{1} \} \cup \ldots \cup
\{N_{2b-1}\geq N_{2b} \} \big]. \nonumber
\end{align}
By applying the union-bound and the subset-inequality ($A \subseteq B$ implies
$\pP(A) \leq \pP(B)$) in the above equation~\cite{durrettp1996}, we get
\begin{align}
\errorprob_n & \leq (2b+1) \max \big\{ \pP(N_0 = 0), \pP(N_0 \geq N_1), \ldots, \nonumber
\\
& \mbox{\hspace{3.8cm}} \pP(N_{2b-1} \geq N_{2b}) \big\} \label{eq:peupperbound}
\\
\mbox{and } \errorprob_n & \geq \max \big\{ \pP(N_0 = 0), \pP(N_0 \geq N_1), \ldots,
\nonumber \\
& \mbox{\hspace{3.8cm}} \pP(N_{2b-1} \geq N_{2b}) \big\}.
\label{eq:pelowerbound} 
\end{align}
From the above equations, the error-exponent in $e_n$ is maximized if the error
exponent of $\max \big\{ \pP(N_0 = 0), \pP(N_0 \geq N_1), \ldots,  \pP(N_{2b-1}
\geq N_{2b}) \big\}$ is maximized. The constant factor $(2b+1)$
in~\eqref{eq:peupperbound} does not contribute to the error-exponent, since it
does not depend on $n$. The error-exponent maximization of the right side
in~\eqref{eq:pelowerbound} is addressed in the next theorem.

\begin{theorem}
Let $s_b = 1/(2b+1)$ and $p_i = \pP(X_i = i s_b)$, for $0 \leq i \leq 2b$, the
chance that a sensor lands at location $i s_b$. Let $N_i = \frac{1}{n} \sum_{i =
0}^{n} \mathbbm{1}(X_i = i s_b)$ be the number of sensors at location $i s_b$.
Then, for $\errorprob_n$ as defined in~\eqref{eq:pe}, the error-exponent is
minimized when the sensor deployment probabilities are chosen according to the
rule
\begin{align}
p^*_{i} = \frac{3 (i+1)^{2}}{(b+1)(2b+1)(4b+3)} \mbox{ for } 0 \leq i \leq 2b.
\end{align}
It can be verified that $\sum_{i = 0}^{2b} p^*_i = 1$ for this rule.
\end{theorem}

\proof This exponent minimizes the detection error probability of the clustering
scheme in Section~\ref{sec:detectalgo}. A sensor falls at location $0$ with
probability $p_{0}$. With $n$ randomly deployed sensors,
\begin{align}
\pP [N_{0}=0] = (1-p_{0})^{n} \label{eq:n0}.
\end{align}
Unlike the first term $\pP[N_0 = 0]$, other events $\pP[N_i \geq N_{i+1}], 0
\leq i \leq 2b$ will be difficult to compute exactly since the random variables
$N_0, N_1, \ldots, N_{2b}$ are dependent. To obtain the error exponents of these
terms, Sanov's theorem will be used (see~\eqref{eq:sanov}).

The empirical distribution of the cluster types is $\vec{q} =
\left[\frac{N_{0}}{n}, \frac{N_{1}}{n}, \ldots, \frac{N_{2b}}{n} \right]$.  In
accordance with the Sanov's theorem, an empirical distribution $\vec{q}$ will be
found such that $D(\vec{q} \parallel \vec{p})$ 	is minimum, subject to the
condition $N_0 > N_1$.  This will result in the error exponent of the event
$\pP(N_0 > N_1)$ (see~\eqref{eq:sanov}).  The empirical distribution is
$ 
\vec{q} = \left[\frac{N_{0}}{n}, \frac{N_{1}}{n}, \ldots, \frac{N_{2b}}{n}
\right] 
$ 
and, from Sanov's theorem, the function to be minimized is
\begin{align}
D(\vec{q} \parallel \vec{p}) & = \sum\limits_{i=0}^{2b}\frac{N_{i}}{n}\log_2
\frac{N_{i}}{np_{i}} \nonumber \\
\mbox{subject to } & \sum\limits_{i=0}^{2b}\frac{N_{i}}{n}=1 \mbox{ and }
N_{1}\leq N_{0}. \label{eq:obj}
\end{align}

The cost-function is convex in the variables $N_i$, and the constraints on $N_i$
are affine; as a result, the optimal solution must satisfy the KKT
conditions~\cite{boydVC2004}. The corresponding Lagrangian is 
\begin{align} 
L=\sum\limits_{i=0}^{2b}\frac{N_{i}}{n}\log_2 \frac{N_{i}}{np_{i}} + \lambda
\left( \sum\limits_{i=0}^{2b} N_i - n \right) + \mu (N_{1}-N_{0} ) \nonumber
\end{align}
At the minima of $D(\vec{q} \parallel \vec{p})$ in~\eqref{eq:obj},
\begin{align}
\frac{\partial L}{\partial N_{i}}=0 \quad \text{for $0\leq i\leq 2b$}
\label{eq:prob}
\end{align}
The solutions of above equation are
\begin{align}
N_{0} = \frac{np_{0}}{e}2^{-n(\lambda - \mu)}, N_{1} =
\frac{np_{1}}{e}2^{-n(\lambda + \mu)}, \label{eq:n0andn1}
\end{align}
and,
\begin{align}
N_{i} = \frac{np_{i}}{e}2^{-n\lambda} \mbox{ for } i \geq 2.  \label{eq:n2up}
\end{align}
According to the KKT~condition for complementary slackness, at the optimal point
we must have~\cite{boydVC2004}
\begin{align}
\mu(N_1-N_0)=0,
\end{align}
which is possible if and only if $\mu=0$ or $N_0=N_1$. If $\mu=0$ then
\begin{align}
N_{i} = \frac{np_{i}}{e}2^{-n\lambda} \mbox{ for all } 0 \leq i \leq 2b. 
\end{align}
Substituting this in the constraint $\sum_{i =0}^{2b} N_i=n$, we get:
\begin{align}
\sum_{i = 0}^{2b} N_{i} = \sum_{i = 0}^{2b} \frac{np_{i}}{e}2^{-n\lambda} = n 
\end{align}
Since $\sum_{i =0}^{2b} p_i = 1$ this reduces to $\frac{2^{-n\lambda}}{e}=1$. Thus the
solution in this case is $N_i=np_i$. However since we only consider the family
of distributions for which $p_0<p_1<\dots <p_{2b}$ this violates the constraint
$N_1\leq N_0$ (since $np_1>np_0$). Hence the optimal solution corresponds to the
case $N_1=N_0$, which gives $\mu= \frac{1}{2n} \log_2 \left( \frac{p_1}{p_0}
\right)$.

For finding $\lambda$, note that $\sum_{i=0}^{2b} N_i = n$. Using $N_0, N_1$
from~\eqref{eq:n0andn1} and $N_i$ from~\eqref{eq:n2up} results in
\begin{align}
\lambda = -\frac{1}{n}\log_2 (\frac{e}{1-(\sqrt{p_{1}}-\sqrt{p_{0}})^{2}})
\end{align}
This value of $\lambda$ gives
\begin{align}
N_{i} &=\frac{np_{i}}{1-(\sqrt{p_{1}}-\sqrt{p_{0}})^{2}} \nonumber \\
\mbox{and } N_{0}=N_{1} & = \frac{n\sqrt{p_{0}p_{1}}}{ 1 -
(\sqrt{p_{1}}-\sqrt{p_{0}})^{2}}. \nonumber
\end{align}
Substitution of $N_0, N_1, \ldots, N_{2b}$ from the above equation
in~\eqref{eq:obj} results in the desired minimum value of $D(\vec{q}*\parallel
\vec{p})$,
\begin{align}
D(\vec{q}*\parallel \vec{p}) =\log_2 \frac{1}{1-(\sqrt{p_{1}}-\sqrt{p_{0}})^{2}}
\end{align}
For $N_i \geq N_{i+1}$, the optimization constraint $N_0 \geq N_1$ will get
replaced by $N_i \geq N_{i+1}$ in~\eqref{eq:obj}. The analysis is identical and
the result is
\begin{align}
D(\vec{q}*\parallel \vec{p}) =\log_2
\frac{1}{1-(\sqrt{p_{i+1}}-\sqrt{p_{i}})^{2}}.
\end{align}
Let $d_{0}=\sqrt{p_{0}}$ and $d_{i}=\sqrt{p_{i}}-\sqrt{p_{i-1}}$, $1\leq i\leq
2b$ and let $d_{\min}=\min\{d_0, d_1, \ldots, d_{2b}\}$.  Then $d_{\min}$ will
determine the value of the largest term in $\max \Big\{ \pP(N_0 = 0), \pP(N_0
\geq N_1), \ldots,  \pP(N_{2b-1} \geq N_{2b}) \Big\}$. This is by Sanov's
theorem which asserts that $\pP(N_i \geq N_{i+1}) \propto 2^{- n D(\vec{q}*
\parallel \vec{p})}$.  Consequently, the value of $d_{\min}$ has to be
maximized.

Let $\vec{p}^*$ be the probability for which $d_{\min}$ is maximized. Then,
\begin{align}
(2b&+1) d_{\min} \leq \sum\limits_{i=0}^{2b}d_{i} = \sqrt{p_{2b}}
\label{eq:dmin}.
\end{align}
To satisfy equality in~\eqref{eq:dmin}, 
\begin{align}
\sqrt{p^*_{0}}=\frac{\sqrt{p^*_{2b}}}{2b+1} \mbox{ and }
\sqrt{p^*_{i+1}}=\sqrt{p^*_{i}}+\frac{\sqrt{p^*_{2b}}}{2b+1}. 
\end{align}
This relationship, along with $p^*_0 +  \ldots + p^*_{2b} = 1$, results in 
\begin{align} 
p^*_{i} = \frac{3 (i+1)^{2}}{(b+1)(2b+1)(4b+3)} \mbox{ for } 0 \leq i \leq 2b.
\label{eq:optdist}
\end{align}
This law on $\vec{p}^*$ ensures that the field detection error probability
in~\eqref{eq:pe} is \textit{minimized}, and is the \textit{main result} of this
work. \endproof

Observe that the probability $e_n$ in~\eqref{eq:pe} will be minimized if $p_0 <
p_1 < \ldots < p_{2b}$ are spaced as far as possible. However, there is a
constraint on the sum of these probabilities. The rule in~\eqref{eq:optdist}
specifies how should elements of $\vec{p}^*$ be spaced to minimize
detection-error probability.

\subsection{Controlling the detection-error probability}

\label{sec:probthresh}

The probability law obtained in the previous section has the minimum detection
error probability that converges to zero asymptotically. It is of interest, in
practical applications where a field is sampled at unknown locations on a
discrete grid, to find the number of samples $n$ which guarantees that any field
of bandwidth $b$ can be estimated with detection-error probability,
$\errorprob_n$, less than some threshold $\epsilon > 0$. 

A sufficient condition for $\errorprob_n \leq \varepsilon$ is
\begin{align}
(2b+1)(1-d_{\min}^{2})^{n}\leq \varepsilon.
\end{align}
Taking logarithm on both sides results in
\begin{align}
n \log_e (1-d_{\min}^{2})&\geq \log_e (\frac{\varepsilon}{(2b+1)})\\
\mbox{or } n &\geq \frac{ \log_e ({\varepsilon}) - \log_e (2b+1)}{\log_e
(1-d_{\min}^{2})}
\end{align} 
This is a sufficient condition on the number of samples required to reduce the
detection error probability below a specified threshold for a field of given
bandwidth.

\subsection{Simulation Results}
\label{sec:simres}

\begin{figure}[!tbp]
  \centering
  \begin{minipage}[b]{0.41\textwidth}
    \includegraphics[width=\textwidth]{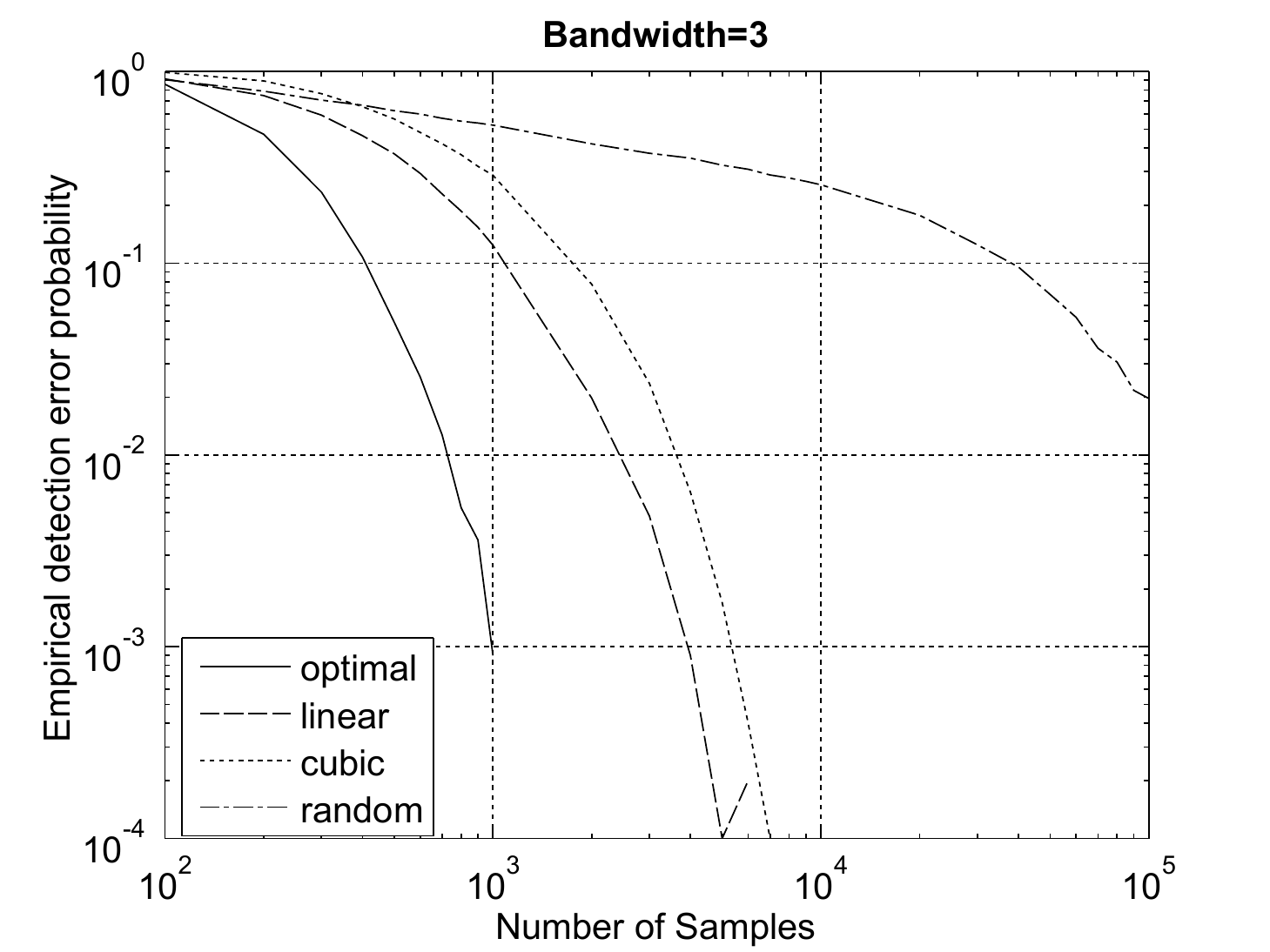}
  \end{minipage}
  \hfill
  \begin{minipage}[b]{0.41\textwidth}
    \includegraphics[width=\textwidth]{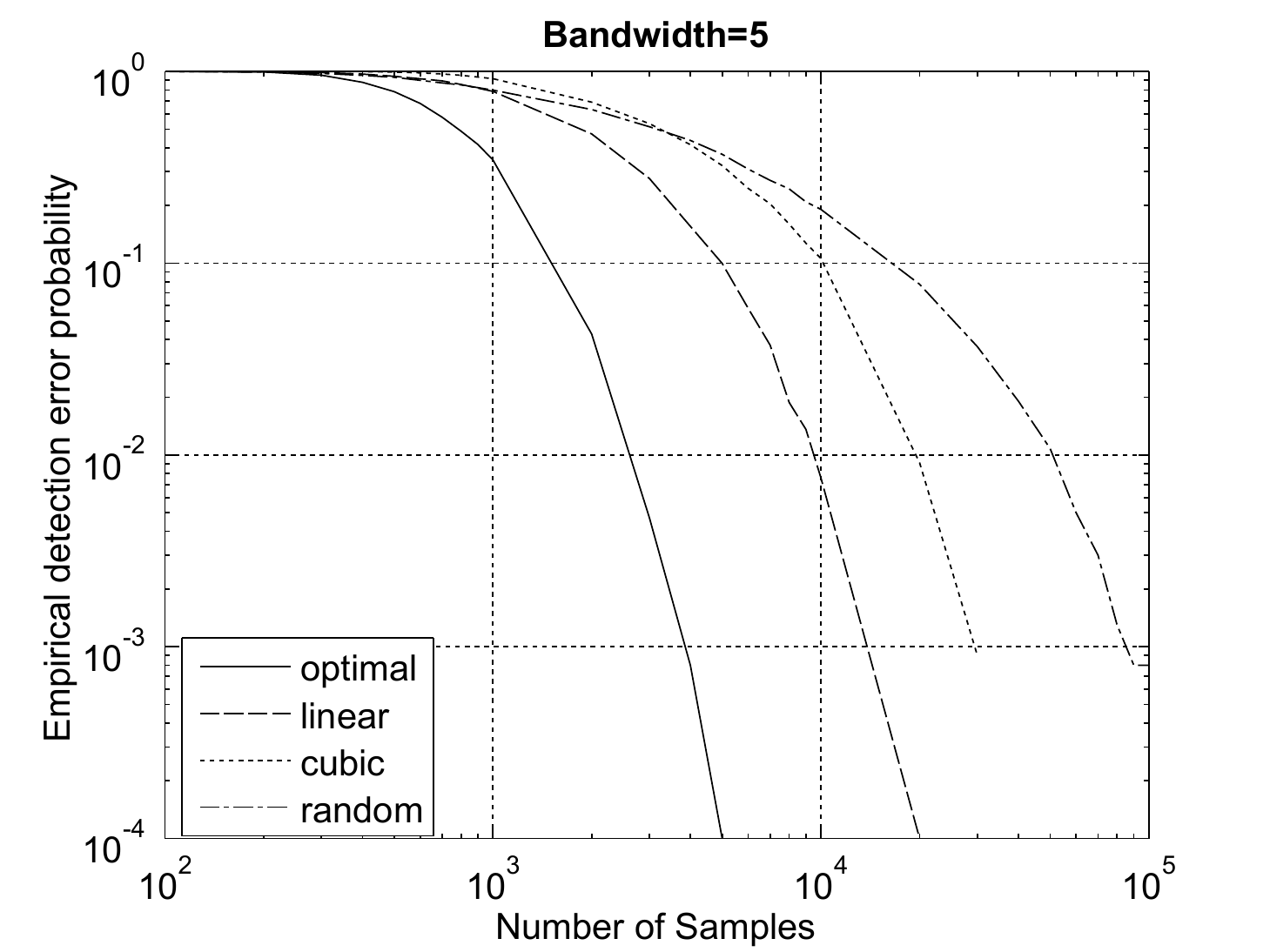}
  \end{minipage}
  \hfill
    \begin{minipage}[b]{0.41\textwidth}
      \includegraphics[width=\textwidth]{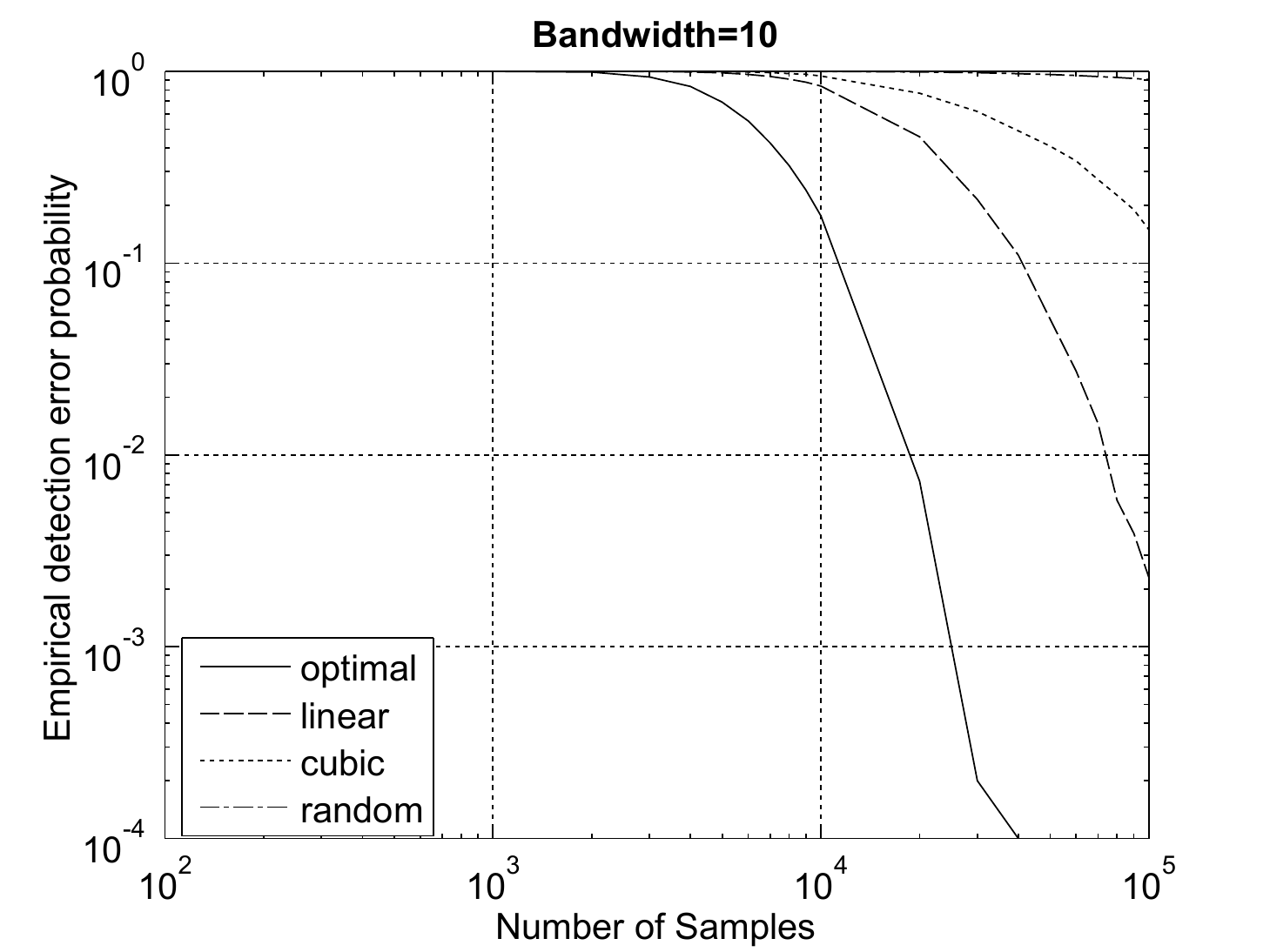}
    \end{minipage}
  \hfill
      \begin{minipage}[b]{0.41\textwidth}
        \includegraphics[width=\textwidth]{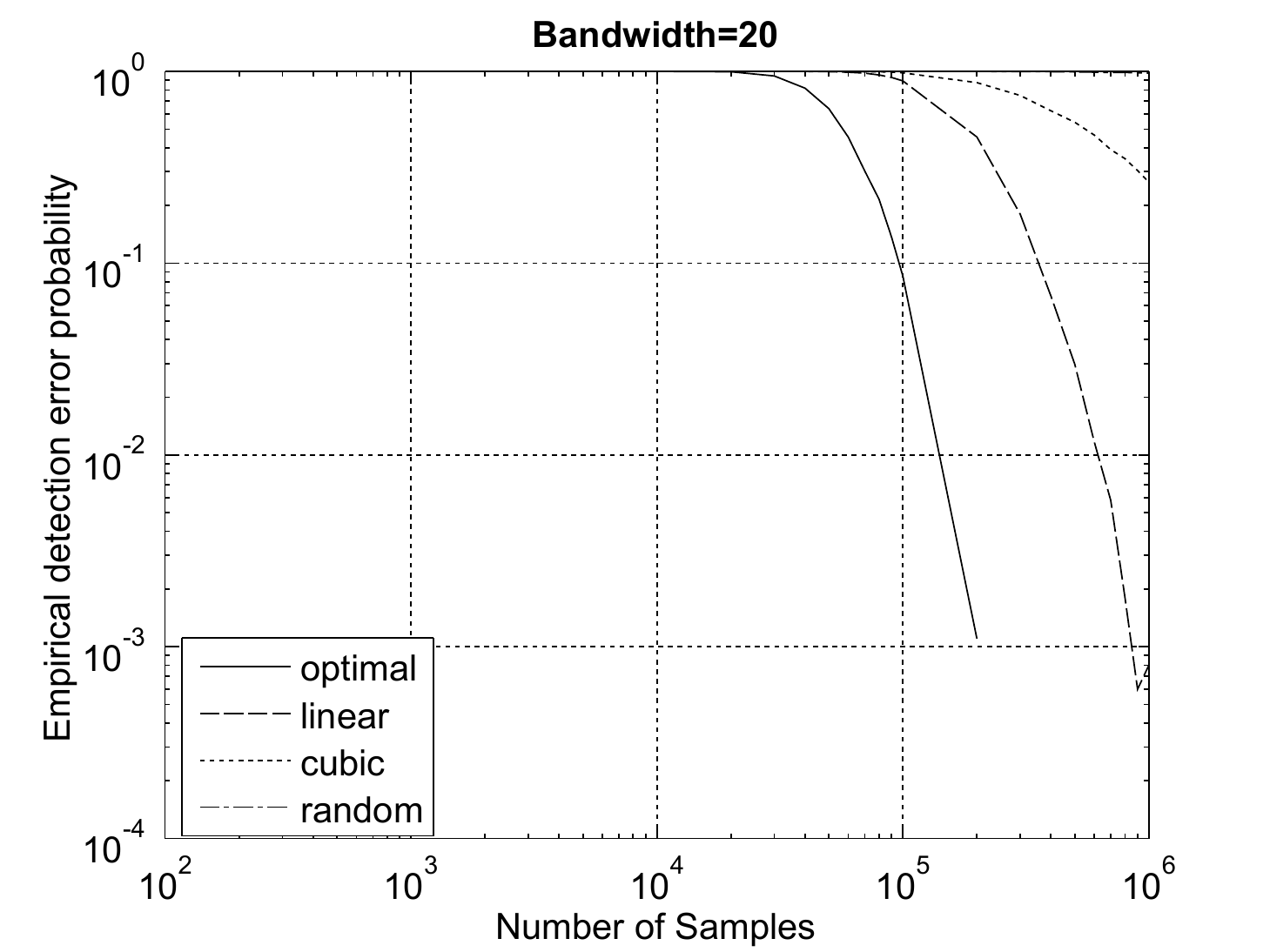}
      \end{minipage}
  \caption{\label{fig:simulations} Detection error-probabilities for different
  laws on $\vec{p}$ and different bandwidths are compared. The four laws used include the optimal $\vec{p}$
  in~\eqref{eq:optdist}, a linear law, a cubic law, and a randomly generated
  $\vec{p}$. Fields of bandwidth $3, 5, 10,$ and $20$ are studied. As expected, the law in~\eqref{eq:optdist} is the best in
  performance in all cases.}
\end{figure}

Using MATLAB, the detection error-probability was compared for different laws on
$\vec{p}$. Fields with bandwidth parameter $b = 3, b = 5, b = 10,$ and $b = 20$
were used.  The real and imaginary parts of the field's Fourier Series
coefficients (for each bandwidth) were selected by i.i.d.~samples of a uniform
random variable. The number of randomly collected samples for each field was
varied between $100$ to $10000$ for the fields of bandwidth $3, 5, 10,$ and
between $100$ to $100000$ for the field of bandwidth $20$. The empirical
detection error-probability, when calculated using $10000$ Monte-Carlo trials,
is plotted in Fig.~\ref{fig:simulations}. A log-log plot is used to understand
the detection error probability exponent.  Four different $\vec{p}$ were used
for comparison and include: (i) the optimal distribution in~\eqref{eq:optdist},
(ii) a linear distribution $\vec{p} = [\alpha_1, 2\alpha_1, \ldots, (2b+1)
\alpha_1]$, (iii) a cubic distribution $\vec{p} = [\alpha_2, 8 \alpha_2, \ldots,
(2b+1)^3 \alpha_2]$, and (iv) ordered uniformly distributed random variable
realizations based distribution $\vec{p} = \alpha_3 [U(1), U(2), \ldots,
U(2b+1)]$.  The constants $\alpha_1, \alpha_2, \alpha_3$ were selected to ensure
that $\sum_{i = 0}^{2b} p_i = 1$.  From the plots, the distribution discovered
in~\eqref{eq:optdist} results in smallest detection error-probability (as
expected) for all bandwidths. The number of samples required to reach zero
detection error probability increases with increasing bandwidth and the optimal
distribution in~\eqref{eq:optdist} is the one whose detection-error probability
decays fastest to zero in all the observed cases.

\begin{figure}
\begin{center}
\includegraphics[width=3.2in]{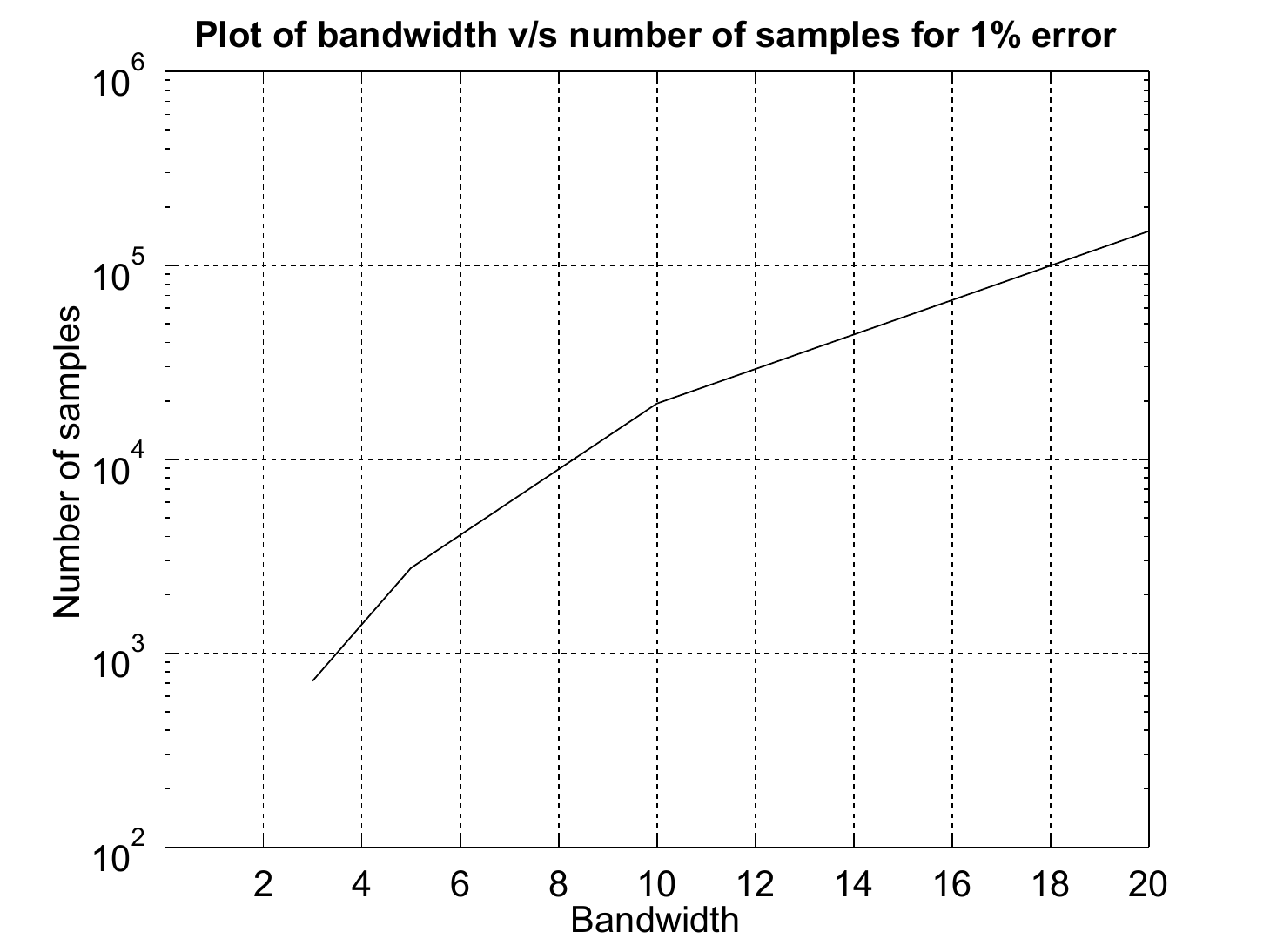} \caption{\label{fig:threshold} Number of
samples required to reduce the empirical detection error probability to $1\%$
for fields of bandwidth $3, 5, 10,$ and $20$}
\end{center}
\end{figure}

For the optimal distribution we also simulated the number of samples required to
reduce the empirical detection error probability $P_{e}$ to $1\%$ for fields of
bandwidth $3, 5, 10,$ and $20$ respectively. The Fourier Series coefficients of
each field was picked by a uniform random number generator. A binary search
algorithm was used to locate the sample size for $0.01-0.001\leq P_{e}\leq
0.01+0.001$. The tolerance of $0.001$ is used since the detection error
probability is calculated as the fraction of incorrectly detected samples from
Monte Carlo simulations and so it need not be exactly equal to $0.01$. The
results are plotted in Fig.~\ref{fig:threshold}.

\section{Reconstruction from Noisy Samples}
\label{sec:noisy}

In this section, the effect of measurement-noise will be evaluated on the
location-unaware sensing scheme analyzed so far.  Consider the case where the
field is sampled as described in Section~\ref{sec:sensordeploy} and the samples
are then affected by zero mean, independent, additive Gaussian noise with  a
known variance $\sigma^2$.  In Section~\ref{sec:performance}, the core idea was
to cluster the observed samples into $(2b+1)$ type-value pairs. If additive
measurement-noise is present, no two samples will be equal with high
probability. In that case, type-value pair based field detection technique will
have to be leveraged by clustering noise-affected samples. This section explores
the clustering of $n$ noise-affected samples into $(2b + 1)$ clusters
quantitatively.

It is assumed that the distribution with which the sensors are deployed at
respective grid-point remains as in~\eqref{eq:optdist}. Since measurement-noise
will affect the field detection process further, it is expected that the
probability of error for field detection in the noisy case will be higher.  This
increase in probability of error in field detection will be characterized in
this section.  Any noise free sample $g(T_i), 1 \leq i \leq n$ has a value equal
to $g(j s_{b})$ for some $0\leq j\leq 2b$ as discussed in
Section~\ref{sec:detectalgo}.  With zero mean independent Gaussian measurement
noise with variance $\sigma^2$, our field detection algorithm is as follows:
\begin{enumerate}
\item The samples $Y_1 := g(T_1)+ W_1, \ldots, Y_n := g(T_n)+ W_n$, with $T_i$
unknown  and in the set  $\{0, s_b, \ldots, 2b s_b\}$, and $W_1, W_2, \ldots,
W_n \sim \mathcal{N}(0,\sigma^{2})$ are obtained.
\item Since $T_{i}=ks_{b}$ with probability $p_{k}$ for all $1\leq i\leq n$, the
readings $Y_{i}$ follow the following Gaussian mixture model (GMM) distribution:
\begin{equation}
f_{Y}(y) = \sum\limits_{k=0}^{2b} p_{k} G(y,g(ks_{b}),\sigma^{2})
\end{equation}
where
\begin{equation}
G(y,g(ks_{b}),\sigma^{2}) = \frac{1}{\sqrt{2\pi} \sigma} \exp\left( 
-\frac{(y-g(ks_{b}))^2}{2\sigma^2} \right) \label{eq:gmm}
\end{equation}

\item Using the readings $Y_1, Y_2, \ldots, Y_n$, the parameters $(p_k, g(k
s_b)), k = 0, 1, \ldots, 2b$ have to be estimated. To address this estimation
problem, the readings $Y_1, Y_2, \ldots, Y_n$ are clustered using the well known
expectation maximization (EM) algorithm~\cite{dempsterLRM1977}, for GMM~parameter
estimation. 

\item The algorithm gives an estimate of the \textit{weights} $p_k$ and
\textit{means} $g(k s_b)$ (analogous to \textit{type} and \textit{value} in the
noiseless case. In the setup, the value of $\sigma$ is assumed to be known.

Let $\widehat{p}_{k}$ and $\widehat{g}(k s_b)$ be the estimated \textit{weights}
and the corresponding \textit{means}, respectively for $0\leq k\leq 2b$.  For
large $n$, the estimated \textit{weight} $\widehat{p}_{k}$ for each
\textit{mean} will be near the correct $p_k$.
\item The \textit{mean} $\widehat{g}(k s_b)$ with smallest \textit{weight} is
estimated as $g(0)$, the \textit{mean} with next smallest \textit{weight} is
estimated to $g(s_b)$, and so on till $g(2b s_b)$.
\end{enumerate}

The EM~algorithm iteratively estimates the parameters of the GMM~by creating a
function for the expectation of the log likelihood function using the current
estimate of parameters (E-step) and maximizing this expected log-likelihood to
compute a new estimate of the parameters (M-step). These two steps are repeated
until the algorithm converges to a maximum of the log likelihood function to
obtain an estimate of the means and the weights.  In this work, EM algorithm for
`soft' segmentation of the data is used~\cite{lauricFS2007}.  The data comprises
of the readings $Y_1, Y_2, \ldots, Y_n$ which are segmented into clusters. A
cluster is defined as the set of all $Y_{i}$ that are obtained from $g(ks_{b})$
for a fixed $k$. Thus there are $2b+1$ clusters in this case. Instead of
assigning each reading to a single cluster, a membership matrix matrix $\Gamma$
is created. The $(i,k)$ element in $\Gamma$ records a value that indicates the
membership of $Y_i$ in the $k^{\mbox{th}}$ cluster.

The membership matrix $\Gamma$ resolves conflicting situations where the values
of any two or more $g(ks_{b})$ are very close in Euclidean distance and their
corresponding clusters overlap. Samples lying in overlapping clusters could have
originated from either of the sampling locations making it difficult to assign
them to a single cluster. 
The algorithm requires an initial guess of the means which is provided using the
$k$-means++ algorithm~\cite{arthurVK2007}.  The weights, $\widehat{p}_{k}$ are
initially assumed to be uniformly distributed.  The variance of the clusters is
known ($\sigma^{2}$) and is fixed at this value. Each iteration of the algorithm
involves the following two steps:

1) The E-Step:
\begin{align}
\gamma_{ik} := (\Gamma)_{i,k} = \frac{G(Y_{i},\widehat{g} (ks_b),\sigma^2)
\widehat{p}_{k}}{\sum\limits_{k=0}^{2b}G(Y_{i}, \widehat{g}(ks_b), \sigma^2)
\widehat{p}_{k}}
\end{align}

2) The M-Step:
\begin{align}
\widehat{g}(k s_b) = \sum_{i=0}^{n} Y_{i} \gamma_{ik} \mbox{ and } 
\widehat{p}_{k} = \frac{\sum\limits_{i=0}^{n}\gamma_{ik}}{n}
\end{align}
The values of $\widehat{g} (k s_b)$ and $\widehat{p}_{k}$ estimated in the M-step are substituted in
the E-step of the next iteration and the process is repeated until the estimates
converge (squared Euclidean distance between the current and previous estimates
of $\widehat{g} (k s_b)$ falls below a specified threshold).

\begin{figure}[h]
  \centering
   \includegraphics[width=3.4in]{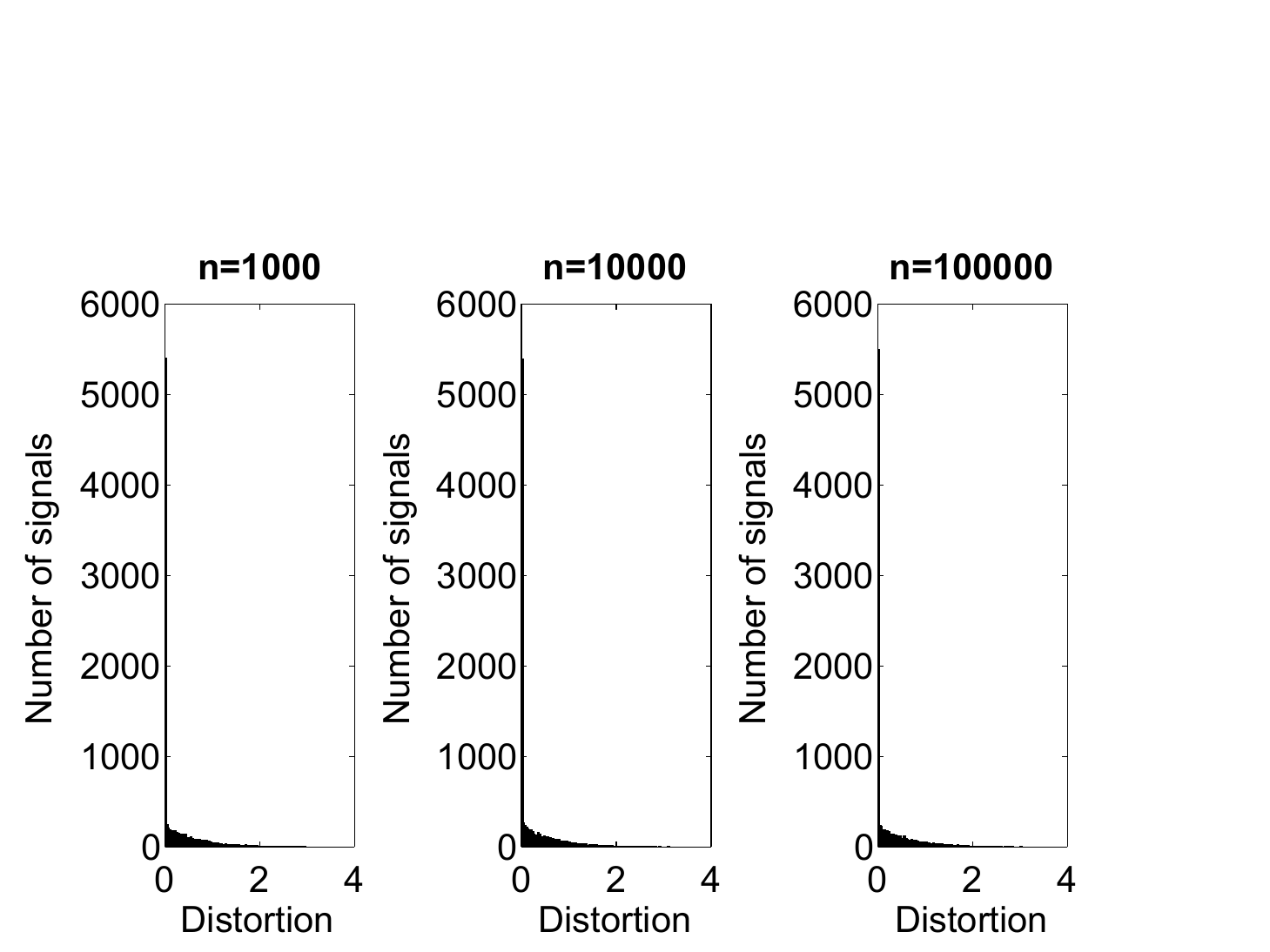}
  \caption{\label{fig:noisy1} Results of the sampling and estimation experiment
for 10000 randomly generated signals of bandwidth parameter $b=3$. Histograms of the distortion are plotted for each sample size ($n$)}
\end{figure}

\begin{figure}[h]
  \centering
   \includegraphics[width=3.4in]{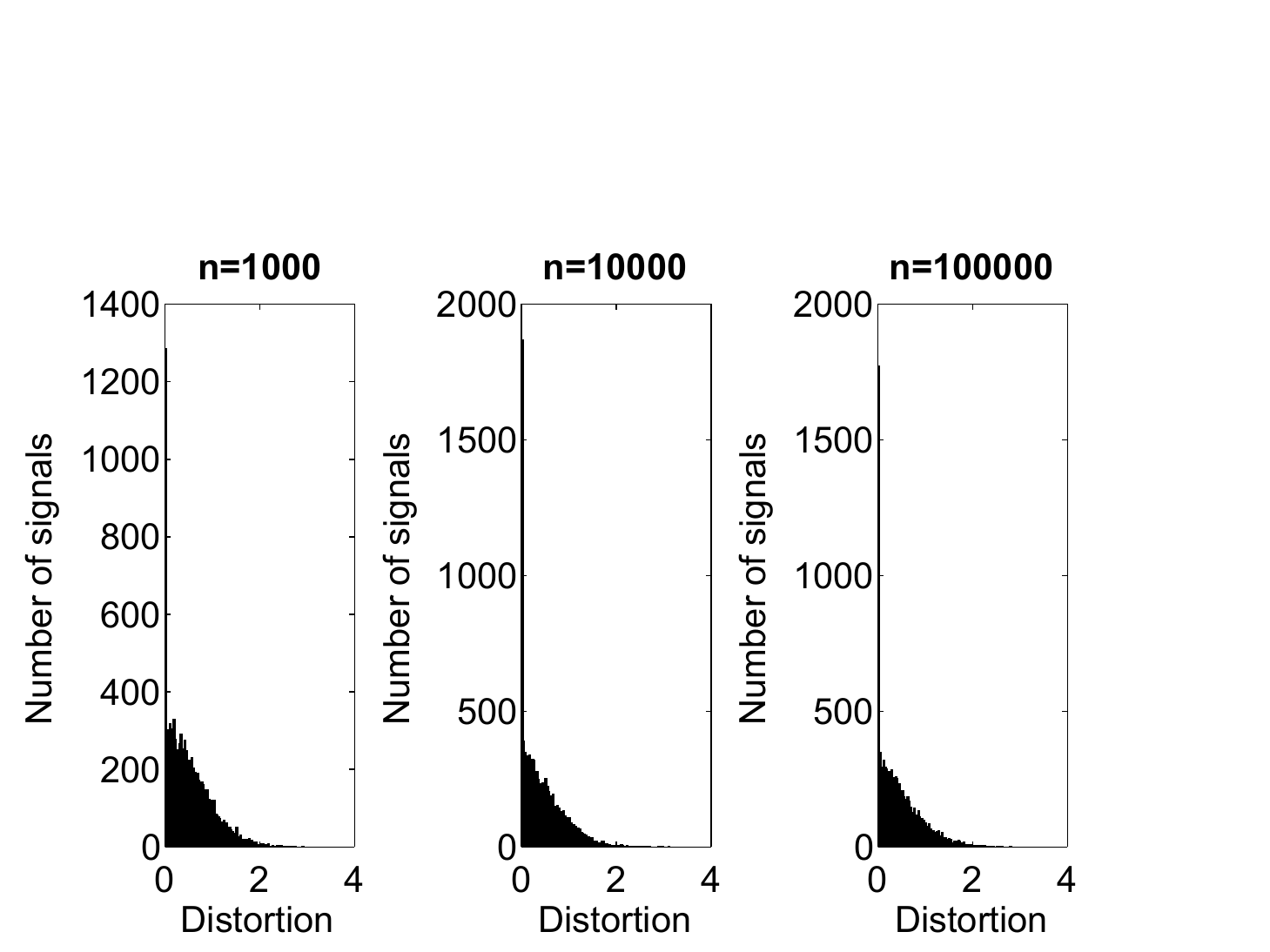}
  \caption{\label{fig:noisy2} Results of the sampling and estimation experiment
for 10000 randomly generated signals of bandwidth parameter $b=5$. Histograms of the distortion are plotted for each sample size ($n$)}
\end{figure}

\begin{figure}[h]
  \centering
   \includegraphics[width=3.4in]{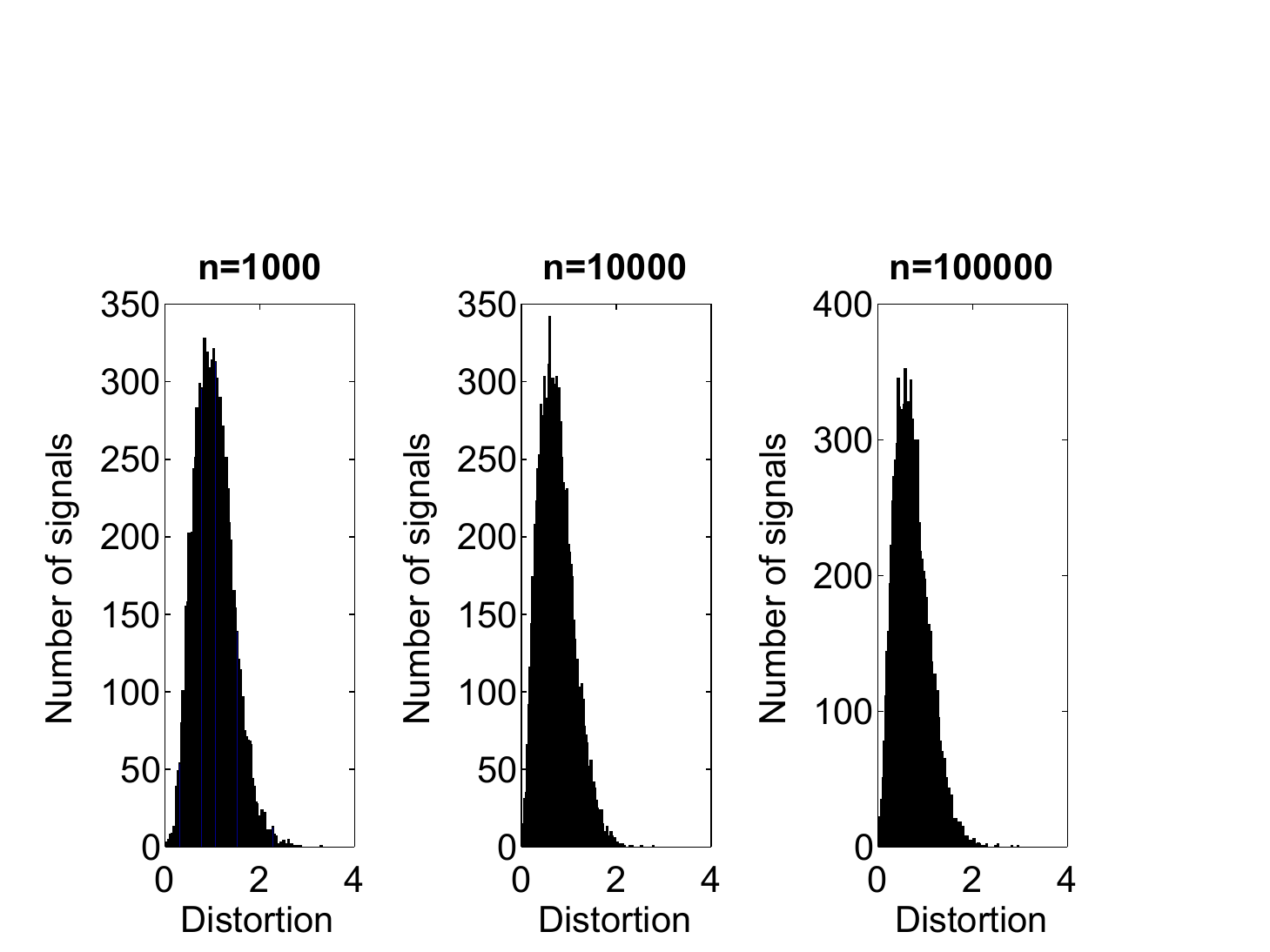}
  \caption{\label{fig:noisy3} Results of the sampling and estimation experiment
for 10000 randomly generated signals of bandwidth parameter $b =10$. Histograms of the distortion are plotted for each sample size ($n$)}
  \end{figure}

The final GMM~estimated by the EM algorithm is:
\begin{equation}
\widehat{f}_{Y}(y)=\sum\limits_{k=0}^{2b} \widehat{p}_{k} G(y, \widehat{g}(k
s_b), \sigma^{2}).
\end{equation}
The above algorithm was simulated using MATLAB on 10000 randomly generated
fields with three values of total number of samples $n=1000,10000,100000$, and
where the samples were corrupted by Gaussian noise ($\mu=0, \sigma=0.05$).  The
experiment was repeated for fields with bandwidth parameter $b = 3, b = 5,$ and
$b = 10$. The following distortion metric was used:
\begin{equation}
D = \frac{\int_0^1 |{\widehat{g}(t) - g(t)} |^{2} \mbox{d}t}{\int_0^1 |g(t)|^2
\mbox{d} t}
\end{equation}
where $D$ is the distortion, $\widehat{g}(t)$ is the estimated field, $g(t)$ is
the original field and the limits of the integral are so chosen because the
field has a period~1. Histograms of the distortion for each of the experiments
are shown in Fig.~\ref{fig:noisy1}, Fig.~\ref{fig:noisy2}, and
Fig.~\ref{fig:noisy3}.

From the histograms in Fig.~\ref{fig:noisy1}, Fig.~\ref{fig:noisy2}, and
Fig.~\ref{fig:noisy3} the performance of the algorithm deteriorates on
increasing the bandwidth $b$. The number of fields reconstructed with a low
value for distortion decreases as the bandwidth changes from $b = 3$ to $b =
10$. More than $50\%$ of the fields are reconstructed with a low value of
distortion for bandwidth $b = 3$ while the number reduces to about $18\%$ for
bandwidth $ b = 5$, and less than $1\%$ for bandwidth $b = 10$.  Increasing the
number of samples drawn improves the performance of the algorithm slightly
especially for the higher bandwidths as illustrated by an increase in the height
of the first bar on the histogram for bandwidths $b = 5$ and $b = 10$ on
increasing the number of samples from $1000$ to $10000$.

\begin{figure}[h]
  \centering
  \includegraphics[width=3.4in]{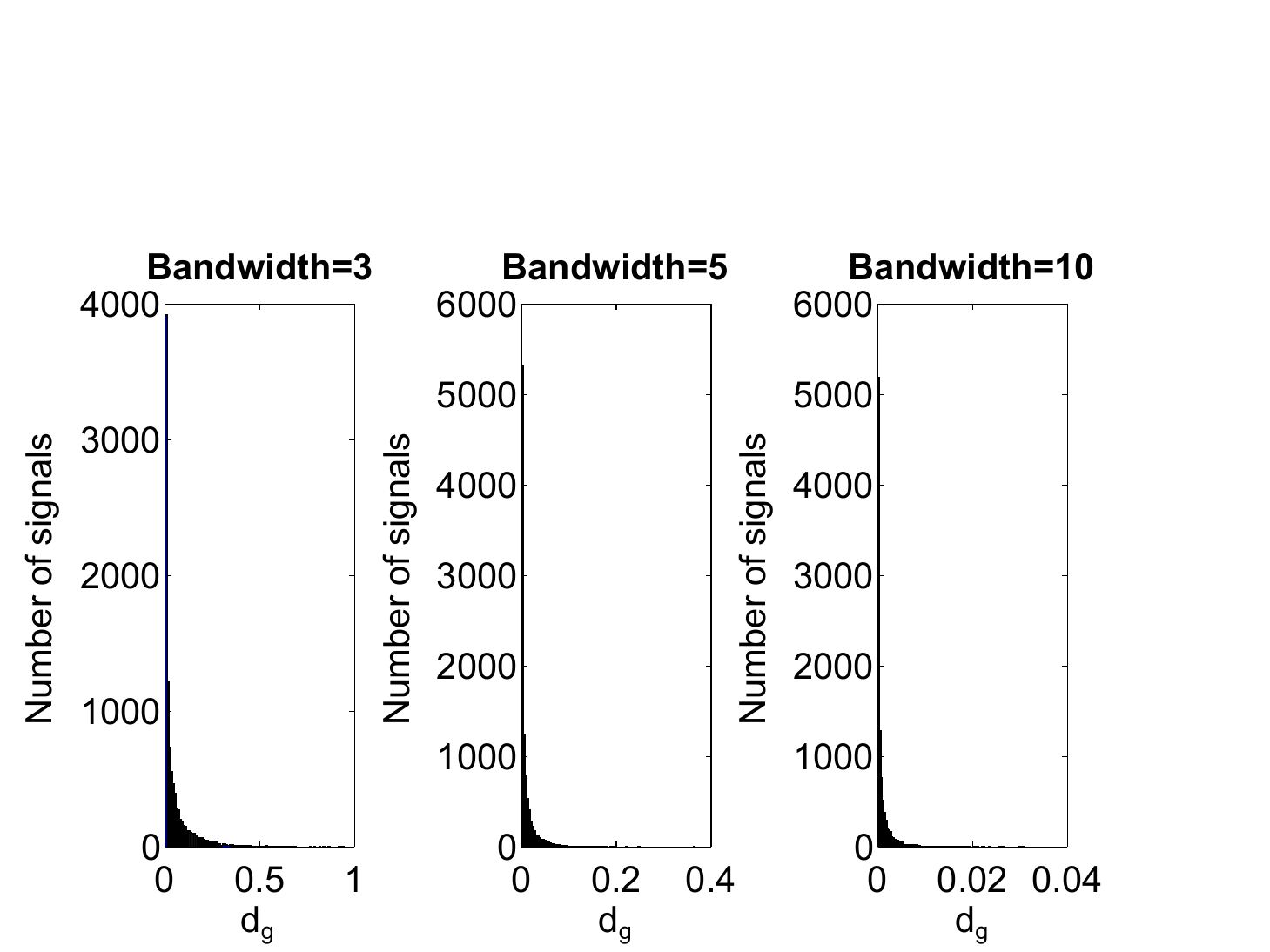}
\caption{\label{fig:dist} Minimum pairwise squared Euclidean distance ($d_{g}$) between the signal values at the sampling locations, is compared for signals of bandwidth $3, 5$ and $10$. Histograms of $d_{g}$ are plotted using $10000$ randomly generated signals for each value of bandwidth}
\end{figure}

For fields with a higher value of distortion it is anticipated that this
high value is due to a large degree of overlap between the clusters. This is
observed in Fig.~\ref{fig:dist}, which is a histogram of the minimum pairwise
squared Euclidean distance between the values of the underlying field at the
sampling locations, $g(ks_{b})$ for the different experiments. These values
serve as the true cluster means and as can be seen from the histogram plots,
increasing the bandwidth increases the number of cases in which two different
clusters' means lie close to each other.  The closeness of cluster means is
indicated by an increase in the height of the first bar of the histogram and a
decrease in the bin width. This closeness of cluster means explains the poor
performance of the reconstruction algorithm with increasing bandwidth.

For a Gaussian distribution, $\mathcal{N}(\mu,\sigma^{2})$, $99.7\%$ of the data
lies within $[\mu-3\sigma,\mu+3\sigma]$. Thus if $x_{1}$ and $x_{2}$ are the
means of two normal distributions with the same standard deviation $\sigma$, and
$x_{2}>x_{1}$ the corresponding clusters of samples drawn overlap with a high
probability, if $x_{2}-3\sigma<x_{1}+3\sigma$ or $(x_{2}-x_{1})^2<6\sigma^{2}$.
For $\sigma=0.05$, $6\sigma^{2}=0.09$. Thus if the squared Euclidean distance
between any two cluster centers ($g(ks_{b})$) falls below this value then the
corresponding clusters overlap with a high probability leading to an incorrect
reconstruction.  The problem of overlapping clusters is a common problem in
clustering especially with the EM algorithm. Several approaches to solving this
problem exist in literature (see, e.g.,~\cite{banerjeeKGBMM2005,fuBM2008}). The
application of these approaches to the present problem remains to be studied.

In all our analysis, the distribution of the noise is assumed to independent and
Gaussian. If the distribution of the noise is non-Gaussian, the mixture-model
and the clustering algorithm will have to be changed to suit the noise
distribution.

\section{Conclusion}
\label{sec:conclusions}

Asymmetric (nonuniform) distributions on location-unaware sensors that enable
bandlimited field inference were studied.  For analytical tractability,
location-unaware sensors on a \textit{discrete grid} were studied.  The key idea
was to use associate the samples with their locations by matching the observed
type (frequency) and the expected type.  Based on this key idea, the main result
of this work was to find the \textit{optimal} probability distribution on sensor
locations that minimizes the detection error-probability of the underlying
spatial field. It was shown that the detection error-probability decreases
exponentially fast in the number of sensors deployed. The proposed sampling
algorithm was also extended to include the case of field reconstruction in the
presence of additive measurement-noise. This was achieved by treating the
distribution of the noisy samples as a mixture model and using clustering to
estimate the mixture model parameters.  Simulations which explored the tradeoffs
between the measurement-noise, increase in bandwidth, and the number of
samples obtained were showcased.

The cases where sensors are located with an arbitrary continuous distribution in
field's support is left for future work.

\bibliographystyle{IEEEbib}
\bibliography{../../../../references}

\end{document}